\documentclass[twocolumn,english,aps, prl, showpacs,multibib]{revtex4-1}
\usepackage[T1]{fontenc}
\usepackage[latin9]{inputenc}
\usepackage{xcolor}
\usepackage{babel}

\usepackage{wasysym}
\usepackage{graphicx}
\usepackage{amssymb}
\PassOptionsToPackage{normalem}{ulem}
\usepackage{ulem}
\usepackage[unicode=true, pdfusetitle,
 bookmarks=true,bookmarksnumbered=false,bookmarksopen=false,
 breaklinks=false,pdfborder={0 0 1},backref=false,colorlinks=false]
 {hyperref}

\makeatletter

\providecolor{lyxadded}{rgb}{0,0,1}
\providecolor{lyxdeleted}{rgb}{1,0,0}

\@ifundefined{textcolor}{}
{%
 \definecolor{BLACK}{gray}{0}
 \definecolor{WHITE}{gray}{1}
 \definecolor{RED}{rgb}{1,0,0}
 \definecolor{GREEN}{rgb}{0,1,0}
 \definecolor{BLUE}{rgb}{0,0,1}
 \definecolor{CYAN}{cmyk}{1,0,0,0}
 \definecolor{MAGENTA}{cmyk}{0,1,0,0}
 \definecolor{YELLOW}{cmyk}{0,0,1,0}
 }

\renewcommand{\fnum@figure}{Fig.~\thefigure}
\usepackage{braket}

\makeatother

\begin{document}

\title{Single-site-resolved measurement of the current statistics in optical
lattices}

\author{Stefan Ke{\ss}ler$^{1}$ and Florian Marquardt$^{1,2}$}

\affiliation{$^{1}$Institute for Theoretical Physics, Universit\"{a}t Erlangen-N\"{u}rnberg,
Staudtstra{\ss}e 7, 91058 Erlangen, Germany \\$^{2}$Max Planck
Institute for the Science of Light, G\"{u}nther-Scharowsky-Straße
1/Bau 24, 91058 Erlangen, Germany}
\begin{abstract}
At present, great effort is spent on the experimental realization
of gauge fields for quantum many-body systems in optical lattices.
At the same time, the single-site-resolved detection of individual
atoms has become a new powerful experimental tool. We discuss a protocol
for the single-site resolved measurement of the current statistics
of quantum many-body systems, which makes use of a bichromatic optical
superlattice and single-site detection. We illustrate the protocol
by a numerical study of the current statistics for interacting bosons
in one and two dimensions and discuss the role of the on-site interactions
for the current pattern and the ground-state symmetry for small two-dimensional
lattices with artificial magnetic fields. 
\end{abstract}

\pacs{67.85.-d, 03.65.Vf, 05.60.Gg }

\maketitle
\emph{Introduction.---}Improved and new detection techniques have
contributed significantly to the rapid progress in the study of strongly
correlated states of ultracold atoms in optical lattices in recent
years. The combination of a time-of-flight expansion followed by absorption
imaging gives access to the momentum distribution and has revealed
the superfluid to Mott insulator transition \citep{Greiner2002}.
Going further, one can study the noise correlations in these absorption
images, e.g., exhibiting the quantum statistics of bosons \citep{Foelling2005}
and fermions \citep{Rom2006}. In addition, the excitation spectrum
of interacting bosons has been measured using momentum-resolved Bragg
spectroscopy \citep{Ernst2010}. However, the most significant step
in recent times has arguably been the introduction of single-site
detection (using fluorescence microscopy) in optical lattices \citep{Bakr2009,Sherson2010}.
This new method has been used to observe the shell structure of a
Mott insulator \citep{Bakr2010,Sherson2010}, the propagation of single
bosons \citep{Weitenberg2011}, a spin impurity \citep{Fukuhara2013},
and the structure of density correlations \citep{Endres2011}. In
the future, it may provide new insights into the buildup of entanglement
in many-body systems and the influence of measurements on quantum
many-body dynamics \citep{Daley2012,Kessler2013,Kessler2012}. 

Here, we discuss a scheme for the spatially resolved measurement of
the current between nearest-neighbor lattice sites. It relies on combining
a bichromatic superlattice and single-site detection. Such a tool
is especially timely in the context of the recent effort towards realizing
gauge fields in optical lattices (for recent reviews, see \citep{Dalibard2011,Lewenstein2012,Goldman2013a}),
with first implementations reported in \citep{Aidelsburger2011,Struck2012,Jimenez-Garcia2012,Aidelsburger2013a,Miyake2013}.
These systems can be used to realize states that support equilibrium
currents. Various methods have by now been proposed to detect interesting
aspects of such systems, e.g., quantum Hall edge states and Chern
numbers in topological insulators, using time-of-flight expansion
\citep{Scarola2007,Alba2011,Zhao2011,Polak2013,Wang2013}, light scattering
\citep{Douglas2011,Goldman2012}, or the time evolution of the real
space density after a quench in the potential \citep{Killi2012,Killi2012b,Goldman2013}.
The protocol discussed below has similarities with the latter approach.
However, in addition to probing the current pattern \citep{Killi2012},
it reveals the full spatially resolved current statistics of the quantum-many
body state, from which correlation functions can be extracted. %
\begin{figure}
\includegraphics[width=1\columnwidth]{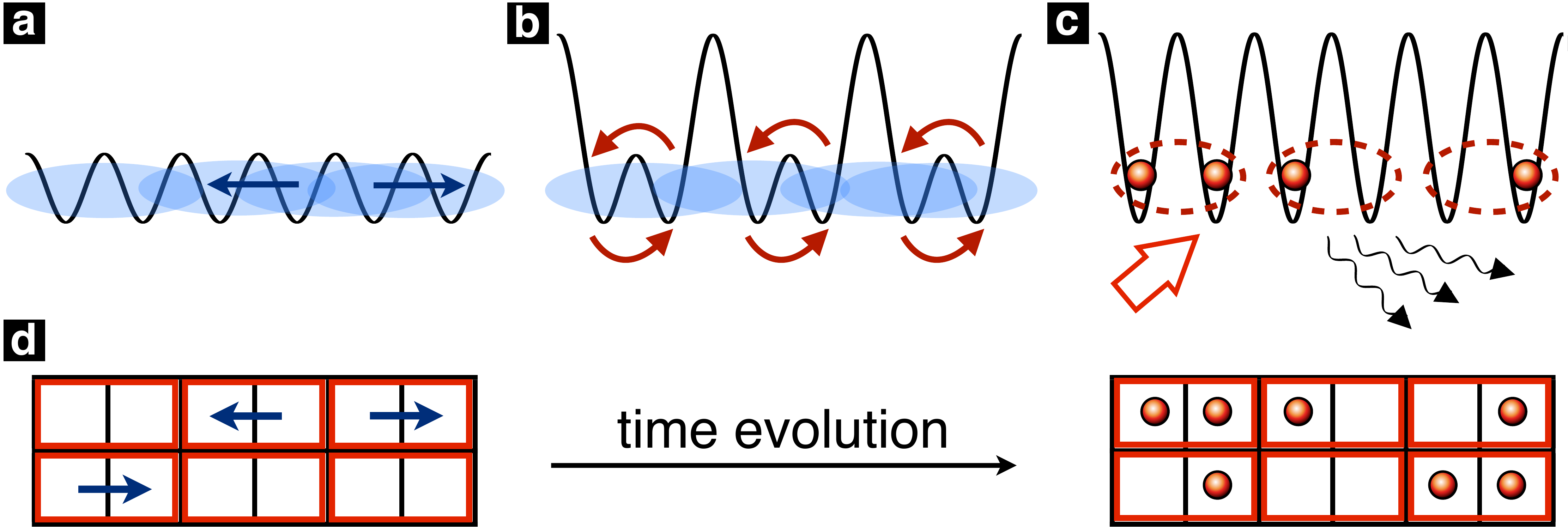}

\caption{(a)-(c) Protocol for measuring the current statistics of a quantum
many-body state in an optical lattice. (b) An additional optical lattice
with double wave length is suddenly ramped up {[}for a two-dimensional
(2D) system, tunneling in the other direction is also turned off{]}
and the on-site interaction between atoms is set close to zero. Afterwards,
the atoms move within a double-well potential. (c) After some time,
the motion is frozen out completely by also ramping up the barrier
within each double well, and the atoms are detected by a single-site-resolved
measurement. (d) An appropriate choice {[}see Eq.~(\ref{eq:current maps density imbalance}){]}
of the evolution time in the double-well potential (b) maps the current
operator to the difference of the particle number at the right and
left lattice site (shown for a 2D setup) and realizes a spatially
resolved measurement of the current operator {[}Eq.~(\ref{eq:Current operator definition}){]}.
\label{fig:Measurement protocol}}

\end{figure}

\emph{Model and current operator.---}We consider a tight-binding Hamiltonian
for ultracold atoms in an optical lattice,\begin{equation}
\hat{\mathcal{H}}=-\sum_{<l,r>}\left\{ J_{lr}\,\hat{c}_{l}^{\dagger}\hat{c}_{r}+J_{lr}^{*}\,\hat{c}_{r}^{\dagger}\hat{c}_{l}\right\} +\sum_{i}\epsilon_{i}\hat{n}_{i}+\hat{\mathcal{H}}_{int}.\label{eq:General Hamiltonian}\end{equation}
Here, $\hat{c}_{l}^{(\dagger)}$ is a bosonic or fermionic annihilation
(creation) operator, $J_{lr}$ is the possibly complex tunneling amplitude,
and $<,>$ denotes pairs of nearest-neighbor lattice sites. The interaction
part $\hat{\mathcal{H}}_{int}$ is a polynomial of local particle
number operators, $\hat{n}_{i}=\hat{c}_{i}^{\dagger}\hat{c}_{i}$,
and $\epsilon_{i}$ is the on-site energy at site $i$. The (mass)
current operator for this system can be defined using the local continuity
equation for the particle density. For a lattice system it reads $\frac{d}{dt}\hat{n}_{l}+\sum_{r\in\mathrm{NN}(l)}\,\hat{\jmath}_{l\rightarrow r}=0$,
here $\hat{\jmath}_{l\rightarrow r}$ denotes the current from site
$l$ to site $r$ and $\mathrm{{NN}}(l)$ denotes the set of nearest-neighbor
lattice sites of $l.$ Using Heisenberg's equation of motion gives
$\frac{d}{dt}\hat{n}_{l}=i\sum_{r\in\mathrm{NN}(l)}\{J_{lr}\,\hat{c}_{l}^{\dagger}\hat{c}_{r}-J_{lr}^{*}\,\hat{c}_{r}^{\dagger}\hat{c}_{l}\}$,
and accordingly the current operator between nearest-neighbor lattice
sites $l$ and $r$ reads (for bosons and fermions)\begin{equation}
\hat{\jmath}_{l\rightarrow r}=-i\left\{ J_{lr}\,\hat{c}_{l}^{\dagger}\hat{c}_{r}-J_{lr}^{*}\,\hat{c}_{r}^{\dagger}\hat{c}_{l}\right\} .\label{eq:Current operator definition}\end{equation}

\emph{Measurement protocol.---}The protocol for measuring the eigenvalues
of the current operator (\ref{eq:Current operator definition}) is
summarized in Fig.~\ref{fig:Measurement protocol}. The main idea
is to use a bichromatic superlattice, which has already been realized
experimentally \citep{SebbyStrabley2006,Trotzky2010}, and to apply
a beam-splitter operation to map the single-particle eigenstates of
the current operator to the states localized at the left and right
lattice site of each double well {[}Fig.~\ref{fig:Measurement protocol}(b){]}.
The measured value of the current operator is then essentially given
by the difference of the particle numbers of both lattice sites. The
particle number can be measured {[}Figs.~\ref{fig:Measurement protocol}(c)
and \ref{fig:Measurement protocol}(d){]} in principle by the recently
developed single-site imaging techniques \citep{Bakr2009,Sherson2010},
although at present these are still restricted to parity measurements
(see the discussion on experimental details below). 

Formally, the protocol relies on the time evolution of noninteracting
atoms in a symmetric double-well potential, $\hat{\mathcal{H}}=-(J_{lr}\hat{c}_{l}^{\dagger}\hat{c}_{r}+J_{lr}^{*}\hat{c}_{r}^{\dagger}\hat{c}_{l}).$
The difference in the atom number at the two wells oscillates in time
and can be expressed as\begin{equation}
\hat{n}_{r}(t)-\hat{n}_{l}(t)=\cos(2Jt)[\hat{n}_{r}(0)-\hat{n}_{l}(0)]+\sin(2Jt)\frac{\hat{\jmath}_{l\rightarrow r}(0)}{J},\label{eq:current maps density imbalance}\end{equation}
where $J=|J_{lr}|.$ Thus, the current can be obtained as the density
difference, $\hat{\jmath}_{l\rightarrow r}(0)=(-1)^{m}J\left[\hat{n}_{r}(\tilde{t})-\hat{n}_{l}(\tilde{t})\right]$,
for suitable chosen evolution times $J\tilde{t}=\pi(2m+1)/4$, $m\in\mathbb{{N}}_{0}$.

The previous expression for $\hat{\jmath}_{l\rightarrow r}(0)$ shows
that the current operator has discrete eigenvalues, just like the
particle number operator. This situation, surprising at first sight,
can also be understood as follows: The eigenvalues of the current
operator are given by the difference in the density of atoms going
to the right and left times the velocity. This is seen by diagonalizing
the current operator (\ref{eq:Current operator definition}), which
yields $\hat{\jmath}_{l\rightarrow r}=J(\hat{c}_{\rightarrow}^{\dagger}\hat{c}_{\rightarrow}-\hat{c}_{\leftarrow}^{\dagger}\hat{c}_{\leftarrow})$
with $\hat{c}_{\rightarrow}=(\hat{c}_{r}+iJ_{lr}^{*}\hat{c}_{l}/J)/\sqrt{2}$
and $\hat{c}_{\leftarrow}=(\hat{c}_{r}-iJ_{lr}^{*}\hat{c}_{l}/J)/\sqrt{2}$.
The operators $\hat{c}_{\rightarrow}$ and $\hat{c}_{\leftarrow}$
have a simple meaning: They correspond to right- and left-going atoms
\citep{Footnote}. Since the total particle number in the double well,
$\hat{n}_{l}+\hat{n}_{r}$, commutes with $\hat{\jmath}_{l\rightarrow r}$,
we can assume a situation of fixed $n_{l}+n_{r}$. Then the spectrum
of the current operator is $J\cdot\{-n,-n+2,\ldots,n\}$, with $n=n_{l}+n_{r}$
for bosons and $n=[n_{l}+n_{r}]\,\mbox{mod}\,2$ for fermions due
to the Pauli principle.

\emph{Bosons in 1D.---}We first consider the current statistics of
the homogeneous one-dimensional (1D) Bose-Hubbard model,\begin{equation}
\hat{\mathcal{H}}=-J\sum_{i}\left\{ \hat{c}_{i+1}^{\dagger}\hat{c}_{i}+\hat{c}_{i}^{\dagger}\hat{c}_{i+1}\right\} +\frac{U}{2}\sum_{i}\hat{n}_{i}(\hat{n}_{i}-1),\end{equation}
with real tunneling amplitude $J$ and on-site interaction strength
$U$. To keep the numerics manageable, we focus on the 1D case, even
though we believe that the qualitative features of the local properties
we are going to discuss should not be dependent on dimensionality
in a significant way. The two phases of the Bose-Hubbard model exhibit
a characteristic atom number statistics at a single site: a Poisson
distribution for the superfluid state ($U/J\rightarrow0$) and a fixed
atom number in the Mott insulating regime ($U/J\rightarrow\infty$)
for integer filling $\bar{n}$; see, e.g., Refs.~\citep{Jaksch1998,Bloch2008}. 

\begin{figure}
\includegraphics[width=1\columnwidth]{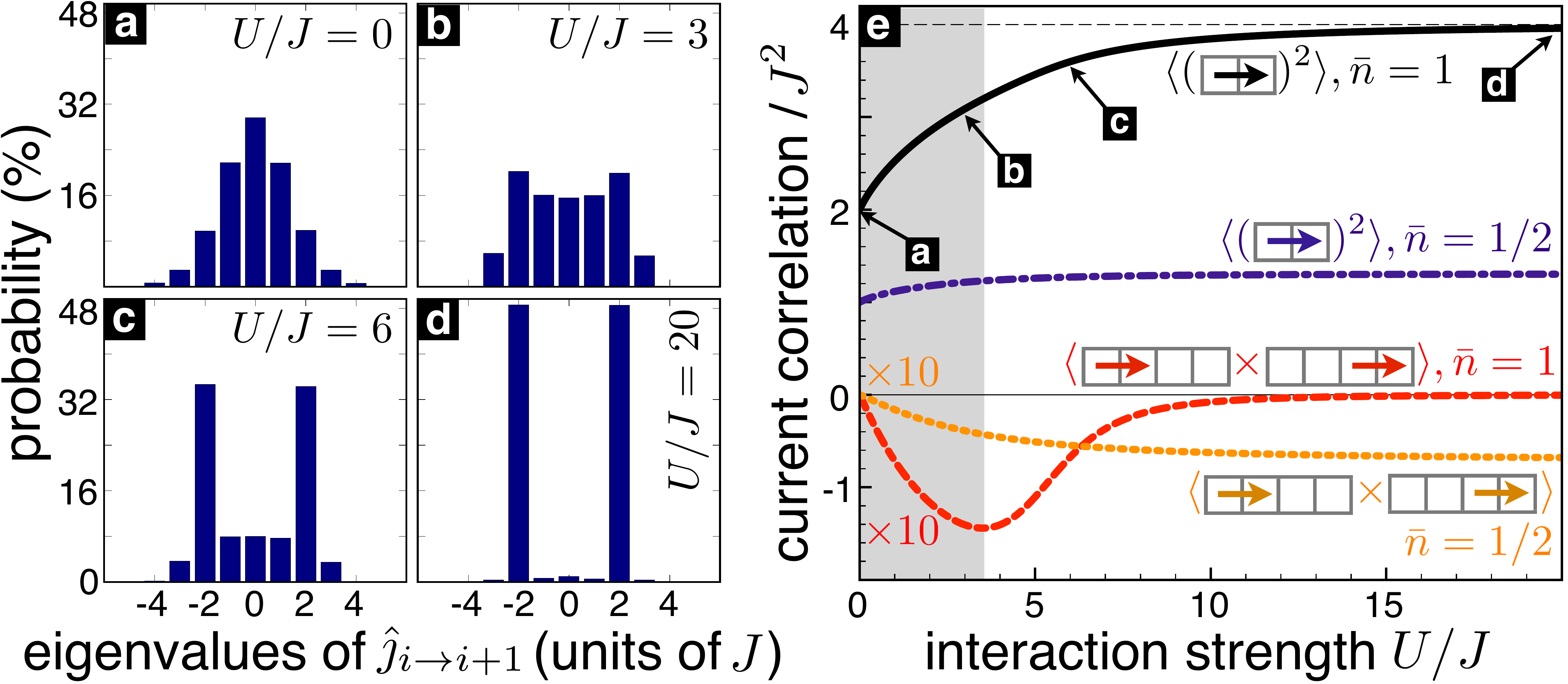}

\caption{Current statistics for the ground state of interacting bosons in a
1D lattice. The results are obtained by exact diagonalization of a
lattice with 12 or 16 sites for filling $\bar{n}=1$ and $\bar{n}=1/2$,
respectively, and periodic boundary conditions. (a)-(d) Distribution
of the current eigenvalues for $\bar{n}=1$ and interaction strength
$U/J=0,3,6,20$ {[}(a)-(d){]} calculated from 25000 snapshots. The
vanishing mean current $\langle\hat{\jmath}_{i\rightarrow i+1}\rangle$
is reflected by the $j\mapsto-j$ symmetry of the distributions. (e)
Interaction dependence of the variance of the current, $\langle(\hat{\jmath}_{i\rightarrow i+1})^{2}\rangle$,
and the current-current correlation $\langle\hat{\jmath}_{i\rightarrow i+1}\hat{\jmath}_{i+2\rightarrow i+3}\rangle$
for unit filling and half filling. The variance of the current increases
monotonically with $U/J$ in both cases. Interestingly, extremal values
of the current-current correlation show up at intermediate interaction
strengths for $\bar{n}=1$. \label{fig:Current correlations 1D system}}

\end{figure}
The corresponding current statistics is presented in Figs.~\ref{fig:Current correlations 1D system}(a)-\ref{fig:Current correlations 1D system}(d)
at filling $\bar{n}=1$. For $U=0$, the atoms are in a product state
of coherent states at the individual sites. Thus, $n_{\rightarrow}$
and $n_{\leftarrow}$ are also Poisson distributed, with mean $\bar{n}$.
The current (\ref{eq:Current operator definition}), being the difference
between two Poisson distributed variables, is then given by the so-called
Skellam distribution, $P_{\bar{n}}(j=Jm)=e^{-2\bar{n}}\mathcal{I}_{|m|}(2\bar{n})$,
where $\mathcal{I}$ denotes the modified Bessel function. When increasing
$U/J$, the distribution becomes more and more concentrated at the
eigenvalues $\pm2J.$ This is a consequence of the Mott insulating
state being a superposition of the eigenstates corresponding to $j=\pm2J$,
$\hat{c}_{i}^{\dagger}\hat{c}_{i+1}^{\dagger}\ket{\mbox{vac}}=\frac{1}{2}[(\hat{c}_{\rightarrow}^{\dagger})^{2}-(\hat{c}_{\leftarrow}^{\dagger})^{2}]\ket{\mbox{vac}}$.
Note that, in general, the current eigenvalues are even multiples
of $J$ for the Mott insulator at arbitrary integer filling.

Figure~\ref{fig:Current correlations 1D system}(e) shows the interaction
dependence of current correlation functions, which might be detected
using the scheme proposed here. The variance of the current increases
monotonically with the on-site interaction strength, from $\langle\hat{\jmath}_{i\rightarrow i+1}^{2}\rangle_{SF}=2J^{2}\bar{n}$
for $U=0$ to $\langle\hat{\jmath}_{i\rightarrow i+1}^{2}\rangle_{MI}=2J^{2}\bar{n}(\bar{n}+1)$
at integer filling $\bar{n}$ deep in the Mott insulating regime.
For $\bar{n}=1,$ the current correlation between neighboring pairs
of lattice sites, $\langle\hat{\jmath}_{i\rightarrow i+1}\hat{\jmath}_{i+2\rightarrow i+3}\rangle$,
becomes negative for intermediate $U/J$ with a minimum close to the
superfluid-to-Mott-insulator transition, while it vanishes for $U\rightarrow0$
and $U/J\rightarrow\infty$. In contrast, for a half-filled lattice,
$\langle\hat{\jmath}_{i\rightarrow i+1}\hat{\jmath}_{i+2\rightarrow i+3}\rangle$
decreases asymptotically as one goes into the hardcore boson limit
($U/J\rightarrow\infty$).

We note that the correlations of the current \emph{into} and \emph{through}
a lattice site might be accessed by an extended version of the measurement
scheme using a triple well-superlattice structure; see Supplemental
Material \citep{Supplement}.

\emph{Bosons in a synthetic magnetic field.---}We consider interacting
bosons subject to a uniform synthetic magnetic field perpendicular
to the 2D lattice. The corresponding Hamiltonian reads, in Landau
gauge,\begin{eqnarray}
\hat{\mathcal{H}} & = & -J\sum_{x,y}\left\{ \hat{c}_{x+1,y}^{\dagger}\hat{c}_{x,y}+\hat{c}_{x,y+1}^{\dagger}\hat{c}_{x,y}e^{i2\pi\alpha x}+\mbox{h.c.}\right\} \nonumber \\
 &  & +\frac{U}{2}\sum_{x,y}\hat{n}_{x,y}(\hat{n}_{x,y}-1).\label{eq:BH Hamiltonian with artificial magnetic field}\end{eqnarray}
Here, $x,y$ are the integer $x$ and $y$ coordinates of the lattice
sites, and the phase $2\pi\alpha$, which a boson picks up when circulating
in an anticlockwise direction around a unit cell, encodes the effect
of the magnetic field. For a charged particle, $\alpha$ would equal
the number of flux quanta per unit cell. The single-particle spectrum
of (\ref{eq:BH Hamiltonian with artificial magnetic field}) is given
by the famous fractal {}``Hofstadter butterfly'' \citep{Hofstadter1976}. 

Below we discuss relatively small 2D lattices, which might be realized
first in experiments (with a suitable superlattice structure dividing
the entire lattice into such small plaquettes as implemented for $2\times2$
lattices in \citep{Aidelsburger2011}). A similar system has been
realized with Bose-Einstein condensates in rotating lattices \citep{Tung2006,Williams2010},
where the {}``Lorentz force'' is replaced by the Coriolis force
\citep{Cooper2008}. The creation and observation of topological states
in this setup have been theoretically studied in \citep{Barberan2006,Hafezi2007,Palmer2008,Umucalilar2008,Umucalilar2010,Grass2012}.
Moreover, a transition between ground states of different rotational
symmetry at discrete rotation frequencies was found \citep{Bhat2006,Bhat2006a},
which leads to a discontinuity in the edge current. Here, we study
the effects of finite on-site interactions on such transitions (these
previous works \citep{Bhat2006,Bhat2006a} discussed only the limit
of hardcore bosons). 

\begin{figure}
\includegraphics[width=1\columnwidth]{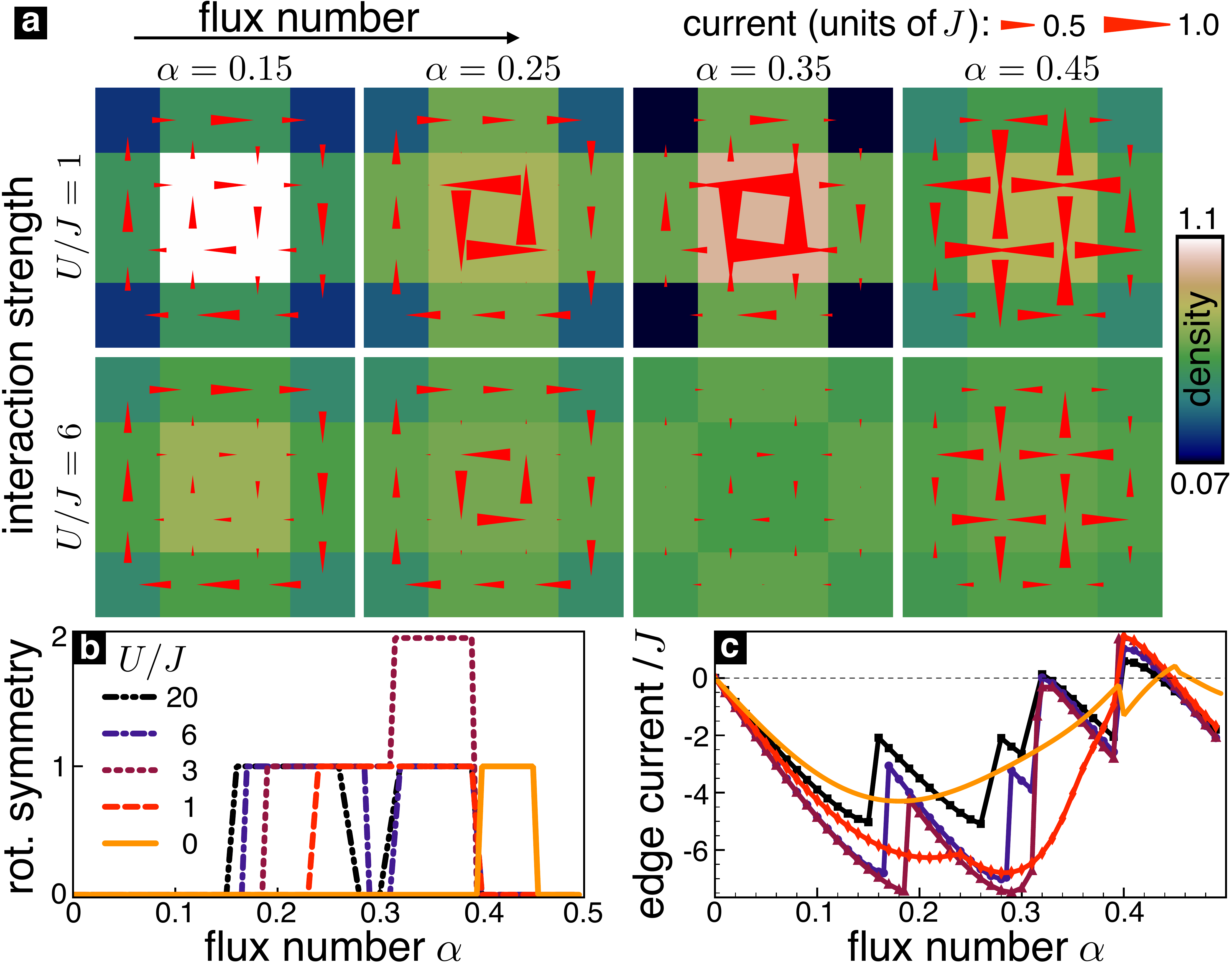}

\caption{Ground-state properties of nine interacting bosons in a $4\times4$
lattice with artificial magnetic field (similar behavior is found
for different fillings): (a) current and density pattern, (b) rotational
symmetry, and (c) edge current. Increasing the flux number starting
from $\alpha=0$, the current flows in clockwise directions and grows
with $\alpha$. The current in the center is generally suppressed
for larger $U/J$, and it reverses its direction above a certain critical
value of $\alpha$ (which decreases with increasing $U/J$). This
change appears together with (b) a transition in the rotational symmetry
of the ground state and (c) a discontinuity in the edge current. Large
interactions lead to additional configurations, as shown, e.g., in
the panel for $U/J=6$ and $\alpha=0.35$. \label{fig:Current correlations for flux lattice}}

\end{figure}

The results obtained from exact diagonalization for a $4\times4$
lattice are summarized in Fig.~\ref{fig:Current correlations for flux lattice}.
Note that we restrict ourselves to the interval $\alpha\in[0,0.5]$,
as the Hamiltonian is invariant under $\alpha\mapsto\alpha+1$, and
$\alpha\mapsto-\alpha$ only changes the magnetic field direction.
Figure~\ref{fig:Current correlations for flux lattice}(a) shows
the current and density profile of the ground state for different
interaction strengths and $\alpha$'s. The current patterns found
for $U/J=1$ are similar to those of $U/J=0$ (not shown), whereas
additional current configurations appear for large $U/J$. Note that
the central current reverses sign beyond some critical flux number
$\alpha$, and this value decreases for larger $U/J$. 

The critical value of $\alpha$ can be identified via the change in
the ground-state rotational symmetry or via the first discontinuity
of the edge current. We define the edge current as the sum of all
currents along the boundary, counted in an anticlockwise direction;
see Fig.~\ref{fig:Current correlations for flux lattice}(c). The
rotational symmetry is best discussed in the symmetric gauge; see
\citep{Supplement}. For a square lattice, the Hamiltonian (\ref{eq:BH Hamiltonian with artificial magnetic field})
commutes with the rotation by $\pi/2$, $\mathcal{R}(\pi/2)$. Thus,
a nondegenerate ground state is an eigenstate of $\mathcal{R}(\pi/2)$,
with eigenvalue $e^{i\pi m/2}$, $m=0,1,2,3$. Additional transitions
in the ground-state symmetry (discontinuities of the edge current)
show up for intermediate $U/J$. The critical flux numbers for these
transition points hardly change for $U/J\apprge10$. The discussed
current patterns and the edge current can be measured using the proposed
protocol, which therefore provides a means of studying the flux and
interaction dependence of such transitions.

\begin{figure}
\includegraphics[width=1\columnwidth]{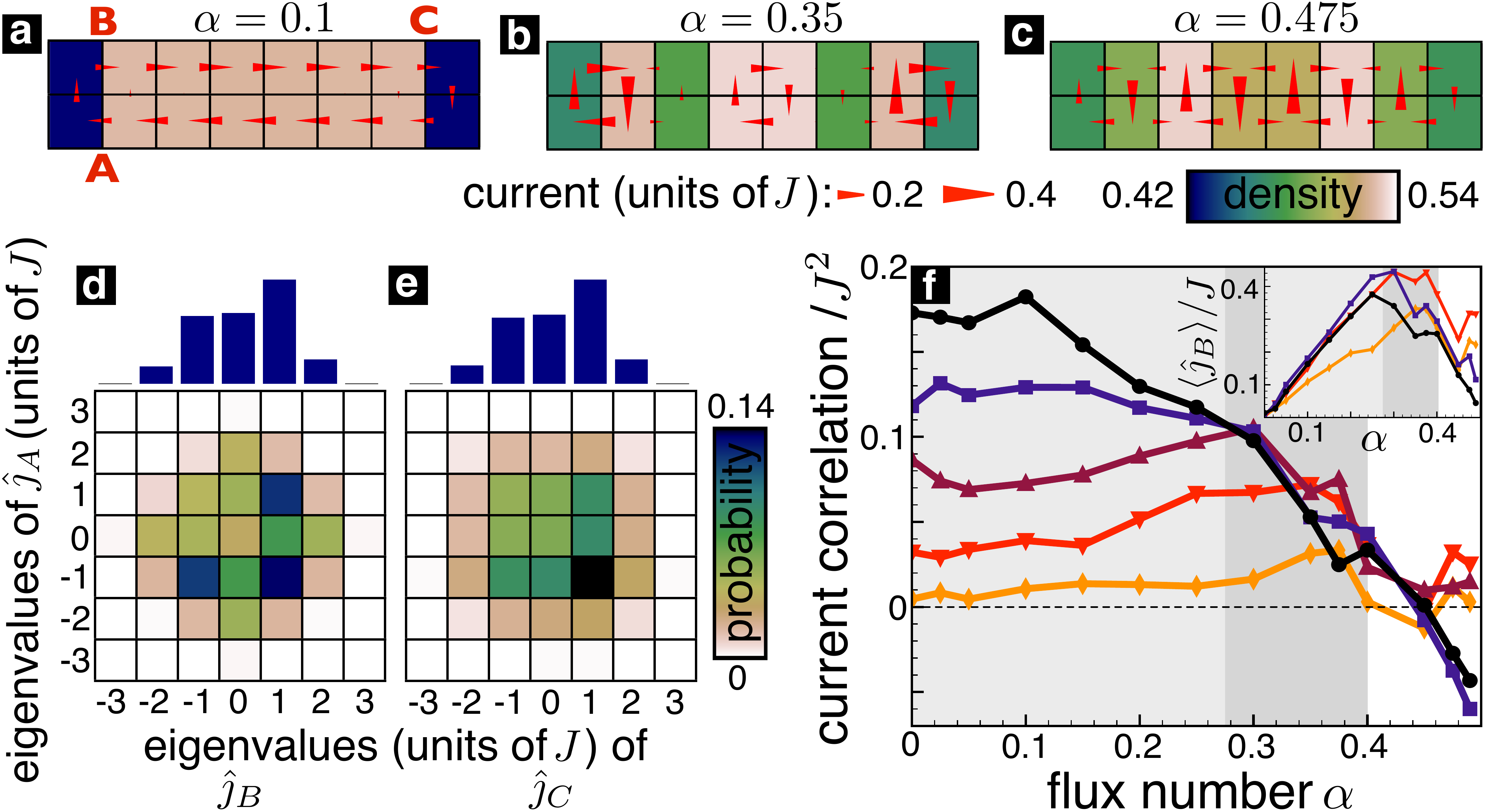}

\caption{Current statistics of the ground state of interacting bosons in a
$8\times2$ lattice at half filling, calculated from 25000 snapshots.
(a)-(c) Density and current profile for interaction strength $U/J=8$
and different flux numbers $\alpha$. (d),(e) Joint eigenvalue distribution
for different current operators and the parameters used in (a). Note
that the currents at the links A,B, and C are defined in the positive
$x$ direction. While $\hat{\jmath}_{B}$ and $\hat{\jmath}_{C}$
have the same eigenvalue distribution (see histograms), their joint
distribution with $\hat{\jmath}_{A}$ is different: $\hat{\jmath}_{A}$
and $\hat{\jmath}_{C}$ are uncorrelated, but $\hat{\jmath}_{A}$
and $\hat{\jmath}_{B}$ are correlated. (f) Current correlation $\langle\hat{\jmath}_{A}\hat{\jmath}_{B}\rangle-\langle\hat{\jmath}_{A}\rangle\langle\hat{\jmath}_{B}\rangle$
for different interaction strengths ($U/J=0.5,2,4,8,25$, bright to
dark). Surprisingly, large on-site interactions lead to a strong positive
current correlation for flux numbers corresponding to the current
pattern (a). The inset shows the $\alpha$ dependence of the mean
current $\langle\hat{\jmath}_{B}\rangle$ for the same interaction
strengths (note that $\langle\hat{\jmath}_{A}\rangle=-\langle\hat{\jmath}_{B}\rangle$).
\label{fig:Current correlations for flux lattice 2}}

\end{figure}

We now turn to spatial current correlations. An illustrative example
is displayed in Fig.~\ref{fig:Current correlations for flux lattice 2},
for a half-filled $8\times2$ lattice. Such bosonic flux ladders exhibit
a transition from a Mei{\ss}ner phase {[}Fig.~\ref{fig:Current correlations for flux lattice 2}(a){]}
to a vortex phase {[}Fig.~\ref{fig:Current correlations for flux lattice 2}(b){]}
when $\alpha$ is increased \citep{Orignac2001,Cha2011}. For the
situation shown in Fig.~\ref{fig:Current correlations for flux lattice 2}(a),
we address the question of whether the currents at A, B, and C are
correlated. This is done by constructing the joint probability distribution
of the eigenvalues from an ensemble of snapshots, since the currents
$\hat{\jmath}_{A},\,\hat{\jmath}_{B},$ and $\hat{\jmath}_{C}$ can
be measured simultaneously. Figures~\ref{fig:Current correlations for flux lattice 2}(d)
and \ref{fig:Current correlations for flux lattice 2}(e) display
two examples of these joint probability distributions. The measured
values of $\hat{\jmath}_{A}$ and $\hat{\jmath}_{C}$ at the two far-removed
links A and C are (to a very good approximation) independent of each
other, i.e., the joint probability distribution is just a product
of the eigenvalue distribution of $\hat{\jmath}_{A}$ and $\hat{\jmath}_{C}$
{[}cf.~Fig.~\ref{fig:Current correlations for flux lattice 2}(e){]}.
In contrast, the current operators $\hat{\jmath}_{A}$ and $\hat{\jmath}_{B}$
for nearby sites are correlated as the joint probabilities $p(j_{A}=-J,j_{B}=-J)$
and $p(j_{A}=-J,j_{B}=0)$ are clearly different even though $p(j_{B}=-J)$
and $p(j_{B}=0)$ are almost equal; see Fig.~\ref{fig:Current correlations for flux lattice 2}(d).
The dependence of the current correlation $\langle\hat{\jmath}_{A}\hat{\jmath}_{B}\rangle-\langle\hat{\jmath}_{A}\rangle\langle\hat{\jmath}_{B}\rangle$
on the flux number $\alpha$ is shown in Fig.~\ref{fig:Current correlations for flux lattice 2}(f).
We find a positive correlation for $\alpha\apprle0.3$ {[}parameter
regime with the current pattern shown in Fig.~\ref{fig:Current correlations for flux lattice 2}(a){]},
which becomes stronger with increasing on-site interaction strength.
In contrast, the average current {[}inset in Fig.~\ref{fig:Current correlations for flux lattice 2}(f){]}
hardly changes with the interaction strength for $U/J\gtrsim2$. For
larger flux values, the correlation falls off to values around zero.

\emph{Discussion of experimental details.---}Let us consider the combination
of the proposed measurement protocol with gauge fields created by
laser-assisted tunneling \citep{Jaksch2003,Gerbier2010,Kolovski2011}
implemented in \citep{Aidelsburger2011,Aidelsburger2013,Miyake2013}.
The required 2D lattice consists of alternating columns with different
on-site energies (and may trap different internal states \citep{Jaksch2003,Gerbier2010}).
Tunneling between different columns is only nonzero when it is driven
by additional light fields (which also imprint the phase on the tunneling
amplitude), while bare tunneling exists within each column. Thus,
the bichromatic superlattice for the current measurement could be
applied in the direction of the columns, while the tunneling between
the columns is inhibited by switching off the driving laser fields.

All of the present experiments on single-site-resolved detection only
resolve the parity of the atom number at any lattice site \citep{Bakr2009,Sherson2010}.
For a single or two coupled 1D chains, considered in Figs.~\ref{fig:Current correlations 1D system}
and \ref{fig:Current correlations for flux lattice 2}, one might
let the atoms expand into another direction before the detection process
(similar to \citep{Weitenberg2011}) to avoid double or higher occupancies.
However, for true 2D configurations, one is currently restricted to
small filling factors, until the parity problem can be circumvented
via alternative approaches. We have further investigated these limitations
of the measurement protocol for the configuration shown in Fig.~\ref{fig:Current correlations for flux lattice 2},
by numerically simulating situations with parity detection only. We
also included residual interactions, $U_{res}$, and timing errors
during the evolution in the double-well potential. We find that the
current pattern and the current correlation can still be observed
in the presence of parity detection, even though the absolute value
of the observed current decreases by up to 25\%. The influence of
the residual interaction is of the order of a few percent for $U_{res}/J\leq1/4$.
A timing error $J\tilde{t}=\pi/4+J\Delta t$ leads to a change in
the current and current correlation of less than $0.02J$ and $0.01J^{2}$,
respectively, for $J\Delta t\leq0.05$. 

\emph{Conclusions.---}We have analyzed a protocol for the site-resolved
measurement of the current operator in optical lattices. Using already
available experimental techniques, it can be employed for interacting
bosons at small filling factors. It can, in principle, be extended
to fermions and possibly also to situations with different species.
Measuring the statistics and spatial structure of currents seems a
promising tool to study the physics of interacting ultracold atoms
subject to gauge fields.

\emph{Acknowledgments.---}We thank the DFG for support in the Emmy-Noether
programme and the SFB/TR 12.

\emph{Note added.---}The Mei{\ss}ner phase in a bosonic flux ladder
{[}cf.~Fig.~\ref{fig:Current correlations for flux lattice 2}(a){]}
was recently observed experimentally by measuring the average edge
currents \citep{Atala2014}.

\bibliographystyle{apsrev4-1}
\bibliography{AllReferences}

\begin{thebibliography}{58}
\bibitem [{Foo()}]{Footnote}%
  \BibitemOpen
  \href@noop {} {}\bibinfo {note} {More precisely, if we were to consider an
  extended 1D tight-binding lattice with nearest-neighbor tunneling amplitude
  $J_{lr}$, these would be the states with maximal velocity $\pm2J$, at the
  momenta $k=\pm\pi/2-\arg(J_{lr})$.}\BibitemShut {Stop}%
\end{thebibliography}

\clearpage

\renewcommand{\theequation}{S.\arabic{equation}}

\renewcommand{\thefigure}{S\arabic{figure}}

\setcounter{equation}{0}

\setcounter{figure}{0}

\onecolumngrid

\section*{Supplemental Material for {}``Single-site-resolved measurement of
the current statistics in optical lattices''}

\begin{center}
Stefan Ke{\ss}ler$^{1}$ and Florian Marquardt$^{1,2}$
\par\end{center}

\begin{center}
\emph{$^{1}$Institute for Theoretical Physics, Universit\"{a}t Erlangen-N\"{u}rnberg,
Staudtstra}{\ss}\emph{e 7, 91058 Erlangen, Germany \\$^{2}$Max
Planck Institute for the Science of Light, G\"{u}nther-Scharowsky-Straße
1/Bau 24, 91058 Erlangen, Germany}
\par\end{center}

\twocolumngrid

\section{Extended Measurement scheme}

In the main text we describe a projective measurement of the current
operator between two nearest-neighbor lattice sites. From the measurement
outcomes one can calculate the expectation values of any sum of current
operators of the form~(\ref{eq:Current operator definition}), as,
for instance, the edge current. Here, we discuss an extended setup
with the double-well potential replaced by a triple-well potential
{[}cf. Fig.~\ref{fig:Measurement protocol}(b){]}. This modification
allows to measure the statistics of the (one-dimensional) current
\emph{through} a lattice site, $\hat{\jmath}_{i-1\rightarrow i}+\hat{\jmath}_{i\rightarrow i+1}$.
The obtained statistics can be used (together with the one of the
{}``usual'' setup) to evaluate, e.g., the variance of the (one-dimensional)
current \emph{into} a lattice site, $\hat{\jmath}_{i-1\rightarrow i}-\hat{\jmath}_{i\rightarrow i+1}$.
This variance cannot be conceived from the statistics of $\hat{\jmath}_{i\rightarrow i+1}$
alone as it involves the expectation value $\langle\hat{\jmath}_{i-1\rightarrow i}\hat{\jmath}_{i\rightarrow i+1}\rangle$.

We assume a laser configuration, which allows to create a triple-well
superlattice structure in one spatial direction, such that the dynamics
within each triple well is described by $\hat{\mathcal{H}}^{(3)}=-J\sum_{l=1}^{2}(\hat{c}_{l}^{\dagger}\hat{c}_{l+1}+\hat{c}_{l+1}^{\dagger}\hat{c}_{l})$.
We restrict to the case of a real tunneling amplitude $J$ assuming
the superlattice to be applied in the direction of the columns, where
no phase is imprinted on the tunneling amplitude (see discussion of
experimental details in the main text). Experimentally, the triple-well
superlattice might be realized by the use of a bichromatic lattice
with an additional laser beam with half the wavelength of the short
lattice, as discussed in \citep{Schlagheck2010}. The diagonalization
of the three-site Hamiltonian yields $\hat{\mathcal{H}}^{(3)}=-2J\sum_{n=1}^{3}\cos(\frac{\pi n}{4})\hat{d}_{n}^{\dagger}\hat{d}_{n}$
with $\hat{d}_{n}=\sum_{l=1}^{3}(U^{\dagger})_{nl}\hat{c}_{l}$ and
$U_{ln}=\frac{1}{\sqrt{2}}\sin(\frac{\pi ln}{4})$. Making use of
the expression for the time-dependent annihilation operator, $\hat{c}_{l}(t)=\sum_{n,s=1}^{3}U_{ln}\exp\{i2J\cos(\frac{n\pi}{4})t\}[U^{\dagger}]_{ns}\hat{c}_{s}(0)$,
the current through lattice site 2, $\hat{\jmath}_{1\rightarrow2}+\hat{\jmath}_{2\rightarrow3}$,
at time zero (ramp up of the triple-well potential) expressed in terms
of the one-particle density matrix after an evolution time $t$ equals\begin{equation}
\hat{\jmath}_{1\rightarrow2}(0)+\hat{\jmath}_{2\rightarrow3}(0)=J\sum_{l,n=1}^{3}\hat{c}_{l}^{\dagger}(t)A_{ln}\hat{c}_{n}(t),\end{equation}
with\begin{equation}
A=\left[\begin{array}{ccc}
-\sqrt{2}\sin(\sqrt{2}Jt) & -i\cos(\sqrt{2}Jt) & 0\\
i\cos(\sqrt{2}Jt) & 0 & -i\cos(\sqrt{2}Jt)\\
0 & i\cos(\sqrt{2}Jt) & \sqrt{2}\sin(\sqrt{2}Jt)\end{array}\right].\end{equation}
At time points $J\tilde{t}=\pi(m+\frac{1}{2})/\sqrt{2}$, $m\in\mathbb{{N}}_{0}$,
only the diagonal terms of the one-particle density matrix contribute
and the current is given by\begin{equation}
\hat{\jmath}_{1\rightarrow2}(0)+\hat{\jmath}_{2\rightarrow3}(0)=(-1)^{m}\sqrt{2}J[\hat{n}_{3}(\tilde{t})-\hat{n}_{1}(\tilde{t})].\label{eq:Current density mapping for triple well potential}\end{equation}
The density difference on the right-hand side can be measured by a
single-site-resolved measurement of the atom number. The meaning of
Eq.~(\ref{eq:Current density mapping for triple well potential})
is rather simple: For the time span $J\tilde{t}=\pi/(2\sqrt{2})$
the single-particle eigenstates corresponding to the current eigenvalues
$-\sqrt{2}J,$ 0, and $\sqrt{2}J$ are mapped on the states localized
at the lattice sites 1,2, and 3, respectively. Note that a symmetric
triple-well potential cannot be used for the direct detection of the
current into lattice site 2, $\hat{\jmath}_{1\rightarrow2}(0)-\hat{\jmath}_{2\rightarrow3}(0)$,
as this current operator and three-site Hamiltonian have the common
eigenstate $\frac{1}{\sqrt{2}}(\hat{c}_{3}^{\dagger}-\hat{c}_{1}^{\dagger})\ket{\mbox{vac}}.$
Therefore, there is never a time for which this difference of current
operators can be expressed as a combination of single-site densities.

\begin{figure}
\includegraphics[width=1\columnwidth]{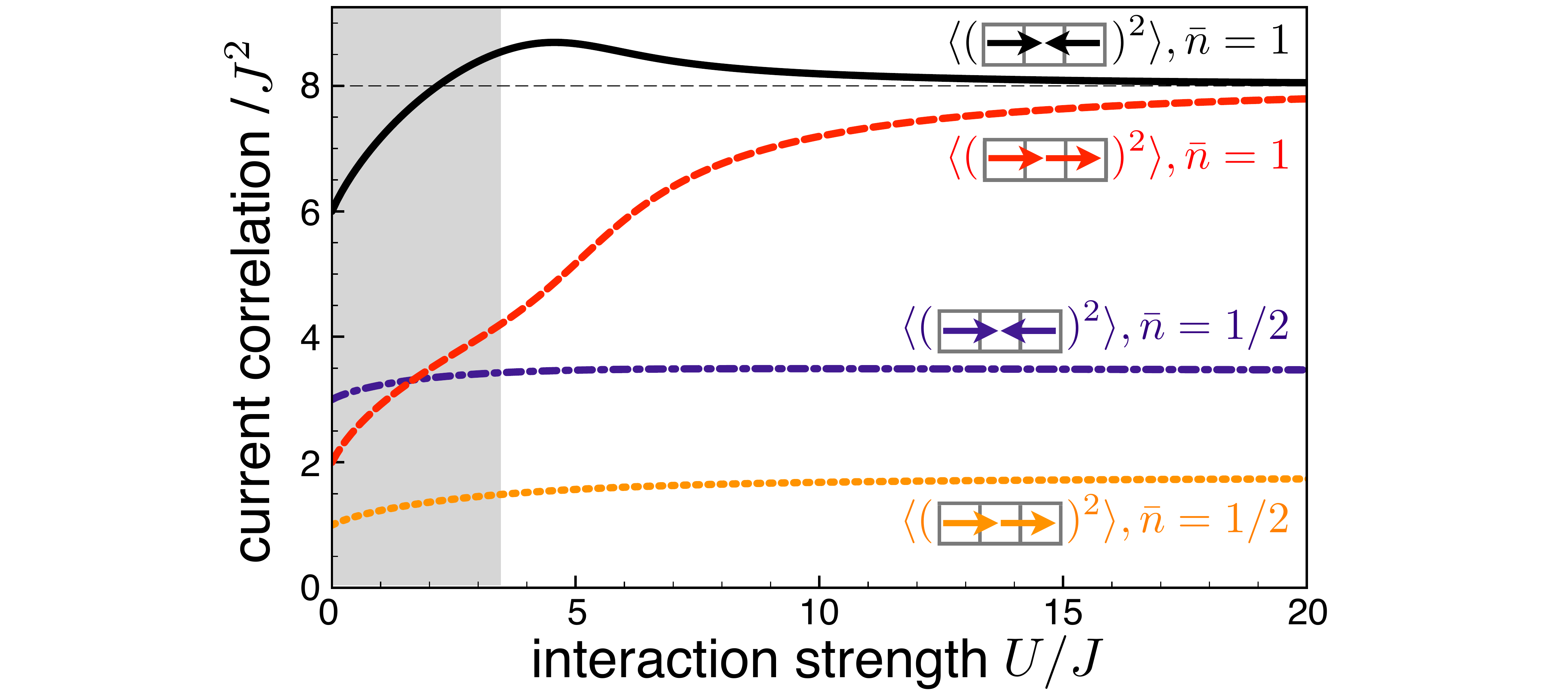}

\caption{Variance of the current \emph{through} and \emph{into} a lattice site
$\langle(\hat{\jmath}_{i-1\rightarrow i}+\hat{\jmath}_{i\rightarrow i+1})^{2}\rangle$
and $\langle(\hat{\jmath}_{i-1\rightarrow i}-\hat{\jmath}_{i\rightarrow i+1})^{2}\rangle$,
respectively, for the ground state of interacting bosons in a 1D lattice.
The parameters are the same as in Fig.~\ref{fig:Current correlations 1D system}.
For integer filling $\bar{n}$, the variances of both currents approach
each other deep in the Mott insulating regime, whereas the fluctuation
in the current \emph{into} a lattice site is much larger than the
fluctuation in the current \emph{through} a lattice site for small
on-site interaction strength $U.$\label{fig:Variance_of_the_current_through_and_into_a_site}}

\end{figure}

Figure~\ref{fig:Variance_of_the_current_through_and_into_a_site}
presents the variance of the current \emph{through} and \emph{into}
a lattice site for the 1D Bose-Hubbard model, which is also considered
in the main text. The mean value of both currents vanishes for the
ground state since $\langle\hat{\jmath}_{i\rightarrow i+1}\rangle=0$;
see also Figs.~2(a-d). Figure~\ref{fig:Variance_of_the_current_through_and_into_a_site}
shows a drastic interaction dependence for a lattice with integer
filling \emph{$\bar{n}.$} For the superfluid state, the fluctuation
of the current \emph{into} a lattice site, and thus $\langle(\frac{d}{dt}\hat{n}_{i})^{2}\rangle$,
is much stronger than the fluctuation of the current \emph{through}
a lattice site. In contrast, they are the same in the Mott insulating
regime in the limit $U/J\rightarrow\infty$.

\section{Rotational symmetry }

Here, we discuss the ground-state rotational symmetry for a many-body
system described by Hamiltonian~(\ref{eq:BH Hamiltonian with artificial magnetic field});
see also \citep{Bhat2006}. We consider a square lattice of $N\times N$
lattice sites {[}for a rectangular lattice the following reasoning
is essentially the same replacing $\hat{R}(\pi/2)$ by $\hat{R}(\pi)$,
which results in two eigenvalues $m=0,1${]}. An anticlockwise rotation
of the coordinate system (passive rotation) by an angle $\pi/2$ maps
the lattice sites onto themselves, $\hat{R}(\pi/2)\ket{x,y}=\ket{y,-x}$,
with the $x$ and $y$ components $x,y\in\{-\frac{N-1}{2},-\frac{N-1}{2}+1,\ldots,\frac{N-1}{2}\}.$
Applying $\hat{R}(\pi/2)$ four times is equivalent to the identity
transformation and thus the possible eigenvalues of $\hat{R}(\pi/2)$
are given by the fourth roots of one, $\lambda_{m}=\exp\{i2\pi m/4\}$
with $m=0,1,2,3.$

Let us now consider the Hamiltonian~(\ref{eq:BH Hamiltonian with artificial magnetic field})
for this square lattice. It reads in the symmetric gauge {[}in Eq.~(5)
we used the Landau gauge{]}, which identifies the center-of-mass of
the lattice as a special point:\begin{eqnarray}
\hat{\mathcal{H}} & = & -J\sum_{y;x\neq(N-1)/2}\left\{ \hat{c}_{x+1,y}^{\dagger}\hat{c}_{x,y}e^{-i\pi\alpha y}+\mbox{h.c.}\right\} \nonumber \\
 &  & -J\sum_{x;y\neq(N-1)/2}\left\{ \hat{c}_{x,y+1}^{\dagger}\hat{c}_{x,y}e^{i\pi\alpha x}+\mbox{h.c.}\right\} \nonumber \\
 &  & +\frac{U}{2}\sum_{x,y}\hat{n}_{x,y}(\hat{n}_{x,y}-1).\label{eq:BH Hamiltonian with artificial magnetic field symmetric gauge}\end{eqnarray}
The transformation of the annihilation (creation) operator under the
rotation, $\hat{R}(\pi/2)\hat{c}_{x,y}^{(\dagger)}\hat{R}(\pi/2)^{T}=\hat{c}_{-y,x}^{(\dagger)}$,
makes it directly apparent that this Hamiltonian commutes with $\hat{R}(\pi/2)$,
i.e., $\hat{\mathcal{H}}=\hat{R}(\pi/2)\hat{\mathcal{H}}\hat{R}(\pi/2)^{T}$.
Thus, for an eigenstate $\ket{\Psi_{i}}$ of $\hat{\mathcal{H}}$,
$\hat{R}(\pi/2)\ket{\Psi_{i}}$ is also an eigenstate of $\hat{\mathcal{H}}$
with the same eigenvalue $E_{i}$. If $E_{i}$ is nondegenerate, then
$\ket{\Psi_{i}}$ is an eigenstate of $\hat{R}(\pi/2)$, too.

In the main text, we discuss the rotational symmetry and the edge
current of the ground state of Hamiltonian~(\ref{eq:BH Hamiltonian with artificial magnetic field})
as function of the parameters $\alpha$ and $U/J$. A change in the
rotational symmetry of a (nondegenerate) ground state happens by an
exact level crossing of the two lowest eigenenergies (as these states
correspond to two different irreducible representations of the rotation;
see \citep{Landau1977}). This implies that the ground-state energy
is nonanalytic at the crossing point and we also observe a discontinuity
in the edge current.

\section{Timing error}

Let us discuss on more general grounds the effect of an imprecisely
chosen evolution time in the double-well potential {[}Fig.~\ref{fig:Measurement protocol}(b){]}
on the current measurement. We consider an evolution time $J\tilde{t}=\pi/4+J\Delta t$,
where $J\Delta t$ is the (dimensionless) timing error. The actually
measured density difference between the left and the right well is
given by Eq.~(\ref{eq:current maps density imbalance}) in the main
text. For a small timing error $J\Delta t\ll1,$ it yields up to second
order in $(J\Delta t)^{2}$:\begin{eqnarray}
\hat{n}_{r}(\tilde{t})-\hat{n}_{l}(\tilde{t}) & = & [1-2(J\Delta t)^{2}]\,\hat{\jmath}_{l\rightarrow r}(0)/J\label{eq:timing error}\\
 &  & -2(J\Delta t)[\hat{n}_{r}(0)-\hat{n}_{l}(0)]+\mathcal{O}([J\Delta t]^{3}).\nonumber \end{eqnarray}
The comparison with the ideal case, $\hat{n}_{r}(\tilde{t})-\hat{n}_{l}(\tilde{t})=\hat{\jmath}_{l\rightarrow r}(0)/J$,
shows that there are two contributions that lead to an error in the
current measurement: one proportional to the true value of the current
and another proportional to the initial density difference between
the two lattice sites. The first term is just a relative change of
the current by $2(J\Delta t)^{2}$, which should be very small in
an experimental realization of the measurement protocol. The second
term seems to be more severe since it is linear in the timing error
$J\Delta t$ and does not depend on the value of the current. However,
for the evaluation of the average current this error is suppressed
for a system with an approximately homogeneous density distribution
or in case that the distribution of timing errors over different measurement
runs is roughly symmetric with respect to $\mbox{\ensuremath{J\Delta t=0.}}$
Indeed, we have found only a small effect of the timing error on the
current and the current correlations for the results shown in Fig.~\ref{fig:Current correlations for flux lattice 2}
(as reported in the main text), where the density distribution is
roughly homogeneous.
\end{document}